\documentclass[aps,prd,preprint,showpacs,superscriptaddress]{revtex4-1}
\usepackage{amsmath}
\usepackage{latexsym}
\usepackage{amsfonts}
\usepackage{graphicx}
\usepackage{epstopdf}

\begin{document}

\newcommand{\tabincell}[2]{\begin{tabular}{@{}#1@{}}#2\end{tabular}}

\title{The Phase-space analysis of scalar fields with non-minimally derivative coupling}

\author{Yumei Huang}
\email{huangymei@gmail.com}
\affiliation{Department of Astronomy, Beijing Normal University, Beijing 100875, China}

\author{Qing Gao}
\email{gaoqing01good@163.com}
\affiliation{MOE Key Laboratory of Fundamental Quantities Measurement, School of Physics, Huazhong University of Science and Technology,
Wuhan, Hubei 430074, China}

\author{Yungui Gong}
\email{yggong@mail.hust.edu.cn}
\affiliation{MOE Key Laboratory of Fundamental Quantities Measurement, School of Physics, Huazhong University of Science and Technology,
Wuhan, Hubei 430074, China}

\begin{abstract}
We perform a dynamical analysis for the exponential scalar field with non-minimally derivative coupling. For the quintessence case,
the stable fixed points are the same with and without the non-minimally derivative coupling. For the phantom case,
the attractor with dark energy domination exists for the minimal coupling only. For the non-minimally derivative
coupling without the standard canonical kinetic term, only the de-Sitter attractor exists, and the dark matter solution
is unstable.
\end{abstract}

\pacs{04.50.Kd, 95.36.+x, 98.80.-k}
\preprint{1412.8152}

\maketitle

\section{Introduction}
Ever since the discovery of the accelerating expansion of the Universe \cite{acc0,hzsst98,scpsn98}, theoretical physicists have
faced the big challenge to explain this phenomenon. A cosmological constant is the simplest way to explain the observed acceleration,
but the theoretical prediction of the cosmological constant is at odds with the observed value by $120$ orders of magnitude.
Furthermore, a cosmological constant also faces the coincidence problem namely why the energy densities of matter and dark energy are nearly equal today.
Dynamical fields with scalar field such as
quintessence \cite{peebles88,wetterich88,quintessence,track1,track2}, phantom \cite{phantom},
tachyon \cite{Sen:2002in,Sen:2002nu,Padmanabhan:2002cp} and k-essence \cite{ArmendarizPicon:2000dh}
were proposed as dynamical dark energy models.
If the accelerating phase is an attractor solution which is independent of initial conditions, then the coincidence problem can be solved.
In particular, the dynamical scalar field has an accelerated scaling attractor and the ratio of the energy densities
between the scalar field and matter is of order 1. For the quintessence model,
exponential potential $V(\phi)=V_0\exp(-\lambda\phi)$ has scaling attractor solutions \cite{Ferreira:1997au,ejcetal98}. For more general scalar fields,
scaling attractor solutions were also found in \cite{copelandde,gong06,Gomes:2013ema}. However, the attractor solution has $\Omega_m=0$
which is inconsistent with the current observation. To solve the coincidence problem with the exponential potential,
phenomenological interactions between dark energy and dark matter were introduced,
but the parameter space was severely constrained \cite{gongplb09,gongjcap09}.

More general models for scalar fields with non-minimal coupling to gravity such as $\xi f(\phi)R$ were also studied extensively.
Recently, a universal attractor behavior for inflation at strong coupling ($\xi\gg 1$) was found for a class of non-minimally
coupled scalar field with the potential $V(\phi)=\lambda^2 f^2(\phi)$.
However, the combination of the non-minimal coupling term $f(\phi)R$ and the Einstein term $R$
can be treated as a special case of the general scalar-tensor theory $F(\phi,R)$.
By a conformal transformation, the non-minimal coupling term $f(\phi)R$ disappears. If the kinetic term of the scalar field
is coupled to curvature, then the model cannot be transformed to scalar-tensor theory by a conformal transformation \cite{Amendola:1993uh}.
In four dimensions, Horndeski derived the most general field equations which are at most of second order in the derivatives of both the metric
$g_{\mu\nu}$ and the scalar field $\phi$ and gave the most general Lagrangian which leads to the most
general second order equations \cite{Horndeski:1974wa}.
In Horndeski theory, the second derivative $\phi_{;\mu\nu}$ is coupled to the Einstein tensor by the
general form $f(\phi,X)G^{\mu\nu}\phi_{;\mu\nu}$, where $X=g^{\mu\nu}\phi_{,\mu}\phi_{,\nu}$. If we only
consider the non-minimal coupling of the scalar field to the curvature which is quadratic in $\phi$ and
linear in $R$, the most
general Lagrangian is \cite{Amendola:1993uh}
\begin{equation}
\label{nmdcl1}
\begin{split}
&L_1=\kappa_1 \phi_{,\mu}\phi^{,\mu}R, \quad L_2=\kappa_2\phi_{,\mu}\phi_{,\nu}R^{\mu\nu},\\
&L_3=\kappa_3 \phi\Box\phi R, \quad \ \, L_4=\kappa_4\phi\phi_{;\mu\nu} R^{\mu\nu},\\
&L_5=\kappa_5 \phi\phi_{,\mu}R^{;\mu}, \quad L_6=\kappa_6\phi^2\Box R.
\end{split}
\end{equation}
Due to the divergencies $(R\phi^{,\mu}\phi)_{;\mu}$, $(R^{\mu\nu}\phi\phi_{,\mu})_{;\mu}$ and $(R^{,\mu}\phi^2)_{;\mu}$,
only $L_1$, $L_2$ and $L_3$ are independent. For a massless scalar field, the non-minimally derivative coupling
$L_1$ and $L_2$ give a de Sitter attractor solution \cite{Capozziello:1999xt,Capozziello:1999uwa}. Furthermore,
the field equations reduce to the second order equations if $\kappa_2=-2\kappa_1=\kappa$ and
the non-minimally derivative coupling becomes $\kappa G^{\mu\nu}\phi_{,\mu}\phi_{,\nu}$ \cite{Sushkov:2009hk}.
Higgs inflation with $\lambda\phi^4$ potential was then discussed with this non-minimal derivative coupling
and it was found that the model does not suffer from dangerous quantum corrections \cite{Germani:2010gm}.
For a massless scalar field without the canonical kinetic term $g^{\mu\nu}\phi_{,\mu}\phi_{,\nu}$, the non-minimally
derivative coupled scalar field behaves as a dark matter \cite{Gao:2010vr,Ghalee:2013smy}. Because of its
rich physics, the non-minimally
derivative coupling $\kappa G^{\mu\nu}\phi_{,\mu}\phi_{,\nu}$ attracted a lot of interest
recently \cite{Daniel:2007kk,Saridakis:2010mf,Sushkov:2012za,Skugoreva:2013ooa,Germani:2010ux,
Germani:2011ua,DeFelice:2011uc,Tsujikawa:2012mk,Sadjadi:2010bz,Sadjadi:2013uza,Minamitsuji:2013ura,Granda:2009fh,
Granda:2010hb,Jinno:2013fka,Ghalee:2014bta,Sami:2012uh,Anabalon:2013oea,Cisterna:2014nua}.
In this paper, we analyze the dynamical evolution of the scalar field with the non-minimal derivative coupling for
an exponential potential.

\section{The dynamics of scalar field with non-minimally derivative coupling}

The action for the non-minimally derivative coupling scalar field is
\begin{equation}
\label{action1}
S=\int d^4x\sqrt{-g}\left[\frac{M_{pl}^2}{2}R-\frac{1}{2}(\epsilon g^{\mu\nu}-\omega^2G^{\mu\nu})\partial_{\mu}\phi\partial_{\nu}\phi-V(\phi)\right]+S_b,
\end{equation}
Where $M_{pl}^2=(8\pi G)^{-1}=\kappa^{-2}$, $S_b$ is the action for the background matter,
the coupling constant $\omega$ has the dimension of inverse mass,
$\epsilon=0$ corresponds to a non-minimally derivative coupling only,
$\epsilon=1$ corresponds to the canonical kinetic term, and $\epsilon=-1$ corresponds to the phantom case.
The energy-momentum tensor for the scalar field is
\begin{equation}
\label{scaltmunu}
\begin{split}
T^\phi_{\mu\nu}=&\epsilon \phi_{,\mu}\phi_{,\nu}-\frac{1}{2}\epsilon g_{\mu\nu}(\phi_{,\alpha})^2-g_{\mu\nu}V(\phi)\\
&-\omega^2\left\{-\frac{1}{2}\phi_{,\mu}\phi_{,\nu}\,R+2\phi_{,\alpha}\nabla_{(\mu}\phi R^\alpha_{\nu)}+\phi^{,\alpha}\phi^{,\beta}R_{\mu\nu\alpha\beta}\right.\\
&+\nabla_\mu\nabla^\alpha\phi\nabla_\nu\nabla_\alpha\phi-\nabla_\mu\nabla_\nu\phi\Box\phi-\frac{1}{2}(\phi_{,\alpha})^2 G_{\mu\nu}\\
&\left. +g_{\mu\nu}\left[-\frac{1}{2}\nabla^\alpha\nabla^\beta\phi\nabla_\alpha\nabla_\beta\phi+\frac{1}{2}(\Box\phi)^2
-\phi_{,\alpha}\phi_{,\beta}\,R^{\alpha\beta}\right]\right\},
\end{split}
\end{equation}
so the energy density and pressure for the scalar field are
\begin{gather}
\label{scale1}
\rho_\phi=\frac{\dot\phi^2}{2}\left(\epsilon+9\omega^2 H^2\right)+V(\phi),\\
\label{scalp1}
p_\phi=\frac{\dot\phi^2}{2}\left[\epsilon-\omega^2\left(2\dot H+3H^2+\frac{4H\ddot\phi}{\dot\phi}\right)\right]-V(\phi).
\end{gather}
when the non-minimally derivative coupling is absent, $\omega=0$, we recover the standard result
\begin{gather}
\label{scale2}
\rho_\phi=\frac{1}{2}\epsilon\dot\phi^2+V(\phi),\\
\label{scalp2}
p_\phi=\frac{1}{2}\epsilon\dot\phi^2-V(\phi).
\end{gather}

By using the flat Friedmann--Robertson--Walker metric, we obtain the cosmological equations from the action (\ref{action1}) and the
energy-momentum tensor (\ref{scaltmunu}) as
\begin{gather}
\label{noderfr1}
3H^2=\kappa^2(\rho_\phi+\rho_b)=\kappa^2\left[\frac{\dot\phi^2}{2}(\epsilon+9\omega^2 H^2)+V(\phi)+\rho_b\right],\\
\label{noderfr2}
\epsilon(\ddot\phi+3H\dot\phi)+3\omega^2[H^2\ddot\phi+2H\dot H\dot\phi+3H^3\dot\phi]+\frac{dV}{d\phi}=0,\\
\label{noderfr3}
2\dot H+3H^2=-\kappa^2(p_\phi+p_b)=-\kappa^2\left\{\frac{\dot\phi^2}{2}
\left[\epsilon-\omega^2\left(2\dot H+3H^2+\frac{4H\ddot\phi}{\dot\phi}\right)\right]-V(\phi)+w_b \rho_b\right\}.
\end{gather}
The background matter energy density is $\rho_b\propto a^{-3(1+w_b)}$ with
constant equation of state $w_b$, it can be dust (including dark matter) with $w_m=0$,
radiation with $w_r=1/3$ or stiff matter with $w_b=1$. For simplicity, we consider the exponential
potential $V(\phi)=\exp(-\lambda \kappa \phi)$ in this work.

In terms of the dimensionless dynamical variables,
\begin{equation}
x=\frac{\kappa\dot\phi}{\sqrt{6}H},\quad
y=\frac{\kappa\sqrt{V}}{\sqrt{3}H},\quad
u=\sqrt{\frac{3}{2}}\omega\kappa\dot\phi,
\end{equation}
the cosmological equations (\ref{noderfr1}) and (\ref{noderfr2}) become
\begin{flalign}
\label{dynaeq1}
&x'=\frac{1}{\sqrt{6}}xt+xs,\\
\label{dynaeq2}
&y'=-\frac{\sqrt{6}}{2}\lambda xy+ys,\\
\label{dynaeq3}
&u'=\frac{1}{\sqrt{6}}ut,
\end{flalign}

where $x'=dx/d\ln a$, the dimensionless
variable $z=\kappa\sqrt{\rho_b/3}/H$ for the background matter density
satisfies the cosmological constraint $\epsilon x^2+y^2+u^2+z^2=1$,
and the auxiliary variables $s=-\dot{H}/H^2$ and $t=\sqrt{6}\ddot\phi/H\dot{\phi}$ satisfy the following relations:
\begin{equation}
\begin{split}
\left(1-\frac{1}{3}u^2+\frac{4u^4}{9\epsilon x^2+3u^2}\right)s=\frac{3}{2}\gamma_b z^2+3\epsilon x^2+3u^2-\frac{2}{\sqrt{6}}
\frac{\lambda x y^2 u^2}{\epsilon x^2+u^2/3},\\
\left(\epsilon x^2+\frac{u^2}{3}\right)t=3\lambda x y^2-3\sqrt{6}\epsilon x^2-\sqrt{6}u^2+\frac{2\sqrt{6}}{3}u^2\,s,
\end{split}
\end{equation}
and $\gamma_b=1+w_b$. In the above system, if $x=u=0$, then the system is not well defined, so we only get those fixed
points with which $x$ and $u$ are not zero at the same time. To get the fixed points with $x=u=0$,
we use the variable $v=(\omega H)^{-1}=3x/u$ to replace the variable $x=uv/3$ for $\omega\neq 0$.
When the kinetic energy is negligible, $\dot\phi=0$, it seems that $x$ should also be zero. Note that $u=0$
does not mean that the scalar field does not evolve, it just means that the scalar field changes very slowly so
that $\dot\phi$ is negligible but not zero. This point can be better understood if we use the dynamical variables
$x$, $y$ and $z$.
The dimensionless energy densities $\Omega_\phi=\epsilon x^2+y^2+u^2=\epsilon u^2 v^2/9+y^2+u^2$ and $\Omega_b=z^2=1-\Omega_\phi$.
The equation of state parameter $w_\phi$ of the scalar field is
\begin{equation}
\label{wphi1}
w_{\phi}=\frac{p_{\phi}}{\rho_{\phi}}=\frac{\epsilon x^2-u^2/3-y^2+2s\,u^2/9-4t\,u^2/9\sqrt{6}}{\epsilon x^2+y^2+u^2}.
\end{equation}
The effective equation of state parameter $w_{eff}$ of the system is
\begin{equation}
w_{eff}=\frac{p_\phi+p_b}{\rho_\phi+\rho_b}=\epsilon x^2-\frac{1}{3}u^2-y^2+\frac{2}{9}s\,u^2-\frac{4}{9\sqrt{6}}t\,u^2+(\gamma_b-1)z^2.
\end{equation}
The deceleration parameter $q=(1+3w_{eff})/2$. If $w_{eff}<-1/3$, then we have accelerating expansion.

It is obvious that the dynamical equations (\ref{dynaeq1})--(\ref{dynaeq3}) consist of an autonomous system.
For the case with $\epsilon=1$ and $\omega=0$, $u=0$ and the system (\ref{dynaeq1})--(\ref{dynaeq3}) reduces
to the quintessence system \cite{Ferreira:1997au,ejcetal98}. For the case with $\epsilon=-1$ and $\omega=0$,
$u=0$ and the system (\ref{dynaeq1})--(\ref{dynaeq3}) reduces
to the phantom system \cite{Hao:2003th}.
For the non-minimally derivative coupling case with $\epsilon=0$, the dynamical analysis was performed in \cite{Gao:2010vr}.
For the case with $\epsilon=1$ and $\rho_b=0$, the dynamical analysis
for a power-law potential was discussed in \cite{Skugoreva:2013ooa}.
By setting $x'=y'=u'=0$ in Eqs. (\ref{dynaeq1})--(\ref{dynaeq3}), we obtain the following critical points.

Point C1 with $(x_{c1},y_{c1},u_{c1})=(0,0,\pm 1)$, it exists when $\omega\neq 0$.
This point corresponds to dark matter solution found in \cite{Gao:2010vr,Ghalee:2013smy} with
the derivative coupled kinetic energy term domination. For this point, we have $\Omega_\phi=1$ and $w_{eff}=w_\phi=0$ and the scalar field
behaves as dark matter even though its potential energy is zero.

Point C2 with $(x_{c2},y_{c2},u_{c2})=(\pm 1/\sqrt{\epsilon},0,0)$, it exists when $\epsilon>0$. For this point, we have
$\Omega_\phi=w_\phi=w_{eff}=1$ and the canonical kinetic energy of the scalar field dominates the energy density, so it behaves like stiff matter.

Point C3 with $(x_{c3},y_{c3},u_{c3})=(x,0,0)$. The existence condition is $\epsilon>0$, $0<\epsilon x^2\le 1$ and $\gamma_b=2$.
For this point, we have $\Omega_\phi=\epsilon x^2$ and $w_{eff}=w_\phi=w_b=1$, only the canonical kinetic energy of the scalar
field contributes to the energy density and the scalar field tracks the stiff matter background.

Point C4 with $(x_{c4},y_{c4},u_{c4})=(0,0,u)$ with $u^2\le 1$, it exists only when $\gamma_b=1$ and $\omega\neq 0$.
For this point, we have $\Omega_\phi=u^2$ and $w_{eff}=w_\phi=w_b=0$. The non-minimally derivative coupling term makes
the only contribution to the energy density of the scalar field and the scalar field tracks the dust background.
The scalar field behaves like dark matter \cite{Gao:2010vr}.

Point C5 with $(x_{c5},y_{c5},u_{c5})=(\frac{\sqrt{6}\gamma_b}{2\lambda},\frac{\sqrt{6\epsilon\gamma_b(2-\gamma_b)}}{2\lambda},0)$.
The existence condition is $\epsilon>0$, $0\le \gamma_b\le 2$ and $\lambda^2>3\epsilon\gamma_b$. It corresponds to the tracking
solution with $\Omega_\phi=3\epsilon\gamma_b/\lambda^2$ and $w_{eff}=w_\phi=w_b$. Since $u_c=0$, the contribution from the
non-minimally derivative coupling is absent, and the result is the same as the quintessence field.

Point C6 with $(x_{c6},y_{c6},u_{c6})=(\frac{\lambda}{\sqrt{6}\epsilon},\sqrt{1-\frac{\lambda^2}{6\epsilon}},0)$.
The existence condition is $\epsilon<0$ or $\epsilon>0$ and $\lambda^2<6\epsilon$. It corresponds to
the scalar field domination solution with $\Omega_\phi=1$ and $w_\phi=w_{eff}=-1+\lambda^2/3\epsilon$.
To get an accelerating solution, we require $\lambda^2<2\epsilon$. Since $u_c=0$, the result is the same
as the quintessence field.

Point C7 with  $(x_{c7},y_{c7},u_{c7},v_{c7})=(0,0,0,0)$, it exists for all the parameters. For $\omega=0$,
this point also exists even though we derived the point under the assumption that $\omega\neq 0$.
This point corresponds to a background matter domination solution with $\Omega_b=1$, $\Omega_\phi=0$ and $w_{eff}=w_b$.

Point C8 with  $(x_{c8},y_{c8},u_{c8},v_{c8})=(0,1,0,0)$, it exists when $\epsilon=0$.
This point corresponds to the effective cosmological
constant solution with $\Omega_\phi=1$ and $w_{eff}=w_\phi=-1$.
The potential energy of the scalar field dominates the energy density.

The fixed points and their existence conditions are summarized in Table \ref{table1}.
For the quintessence case with $\epsilon=1$ and $\omega=0$,
$u=0$ and only the critical points C2, C3, and C5--C7 exist,
but only the points C2 and C5--C7 were found in \cite{Ferreira:1997au,ejcetal98}.
For the non-minimally derivative coupling case with $\epsilon=0$ and $\omega\neq 0$,
only the critical points C1, C4, C7, and C8 present.
The dynamical analysis for this case was performed in \cite{Gao:2010vr},
but only the critical points C4, C7, and C8 were found.

\begin{table}[htp]
\begin{center}
\begin{tabular}{|c|c|c|c|c|c|c|}
\hline
Points & $\Omega_{\phi}$ & $w_{\phi}$ & $w_{eff}$ & Existence & Stability & Acceleration \\ \hline
C1 & 1 & 0 & 0 &  $\omega\neq 0$ & unstable & No\\\hline
C2 &  1 & 1 & 1 & $\epsilon>0$ &unstable & No\\ \hline
C3 &  $\epsilon x^2$ & 1 & 1 & \tabincell{c}{$\epsilon>0$, $\epsilon x^2< 1$\\ and $\gamma_b =2$}  & \tabincell{c}{Stable for $\epsilon>0$,\\$x>0$ and $\lambda^2\ge 6/x^2$} & No\\ \hline
C4 & $u^2$ & 0 & 0 & \tabincell{c}{$u^2< 1$, $\gamma_b =1$ \\and $\omega\neq 0$} & unstable & No\\ \hline
C5 & $\frac{3\epsilon\gamma_b}{\lambda^2}$ & $w_b$ & $w_b$ & \tabincell{c}{$\epsilon>0$, $0<\gamma_b <2$\\ and $\lambda^2>3\epsilon\gamma_b $} &
 \tabincell{c}{Stable for $0<\gamma_b< 2$,\\$\epsilon>0$ and $\lambda^2>3\epsilon\gamma_b$} & No \\ \hline
C6 & 1 & $-1 +\frac{\lambda^2}{3\epsilon}$ & $-1 +\frac{\lambda^2}{3\epsilon}$ & \tabincell{c}{$\epsilon>0$, $\lambda^2<6\epsilon$\\
$\epsilon<0$, all $\gamma_b$, $\omega$ and $\lambda$} &
\tabincell{c}{Stable for $\lambda^2<3\epsilon\gamma_b$ and \\
$\epsilon>0$, or $\epsilon<0$ and $\omega=0$} & \tabincell{c}{Yes if \\ $\frac{\lambda^2}{\epsilon}<2$}
\\ \hline
C7 & 0 & Undefined & $w_b$ & All $\gamma_b$, $\epsilon$, $\omega$ and $\lambda$ & unstable & No \\ \hline
C8 & 1 & -1 & -1 & $\epsilon=0$ & Stable for $\epsilon=0$ & Yes\\\hline
\end{tabular}
\caption{\label{table1} The properties of the critical points. C1: $(x_{c1},y_{c1},u_{c1})=(0,0,\pm 1)$;
C2: $(x_{c2},y_{c2},u_{c2})=(\pm 1/\sqrt{\epsilon},0,0)$;
C3: $(x_{c3},y_{c3},u_{c3})=(x,0,0)$;
C4: $(x_{c4},y_{c4},u_{c4})=(0,0,u)$;
C5: $(x_{c5},y_{c5},u_{c5})=(\frac{\sqrt{6}\gamma_b}{2\lambda},\frac{\sqrt{6\epsilon\gamma_b(2-\gamma_b)}}{2\lambda},0)$;
C6: $(x_{c6},y_{c6},u_{c6})=(\frac{\lambda}{\sqrt{6}\epsilon},\sqrt{1-\frac{\lambda^2}{6\epsilon}},0)$;
C7: $(y_{c7},u_{c7},v_{c7})=(0,0,0)$;
C8: $(y_{c8},u_{c8},v_{c8})=(1,0,0)$.}
\end{center}
\end{table}

To discuss the stability of the autonomous system
\begin{equation}
\label{eomscol}
\textbf{X}'=\textbf{f(X)},
\end{equation}
we need to expand (\ref{eomscol}) around  the critical point $\bf{X_c}$
by setting $\bf{X}=\bf{X_c}+\bf{U}$ with $\textbf{U}$ the perturbations
of the variables considered as a column vector. Thus, for each
critical point we expand the equations for the perturbations up to
the first order in $\textbf{U}$ as
\begin{equation}
\label{perturbation}
\textbf{U}'={\bf{\Xi}}\cdot \textbf{U},
\end{equation}
where the matrix ${\bf {\Xi}}$ contains the coefficients of the
perturbation equations. If the real parts of
the eigenvalues of the matrix ${\bf {\Xi}}$ are all negative, then
the fixed point is a stable point. Applying this procedure, we find the eigenvalues of the
matrix ${\bf {\Xi}}$ and present the stability conditions
for the above critical points C1--C8.

For the point C1, the eigenvalues are $\lambda_1=\lambda_2=3/2$ and $\lambda_3=3-3\gamma_b$, so it is an unstable point.

For the point C2, the eigenvalues are $\lambda_1=6-3\gamma_b $,
$\lambda_2=3\mp\sqrt{\frac{3}{2}}\lambda/\sqrt{\epsilon}$ and $\lambda_3=-3$.
For the case $x_{c2}=1/\sqrt{\epsilon}$, $\lambda_2<0$ when $\lambda>\sqrt{6\epsilon}$.
For the case $x_{c2}=-1/\sqrt{\epsilon}$, $\lambda_2<0$ when $\lambda<-\sqrt{6\epsilon}$.
Since $0\le \gamma_b\le 2$, $\lambda_1\ge 0$ and the point is an unstable point.

\begin{figure}[htp]
\centerline{\includegraphics[width=0.8\textwidth]{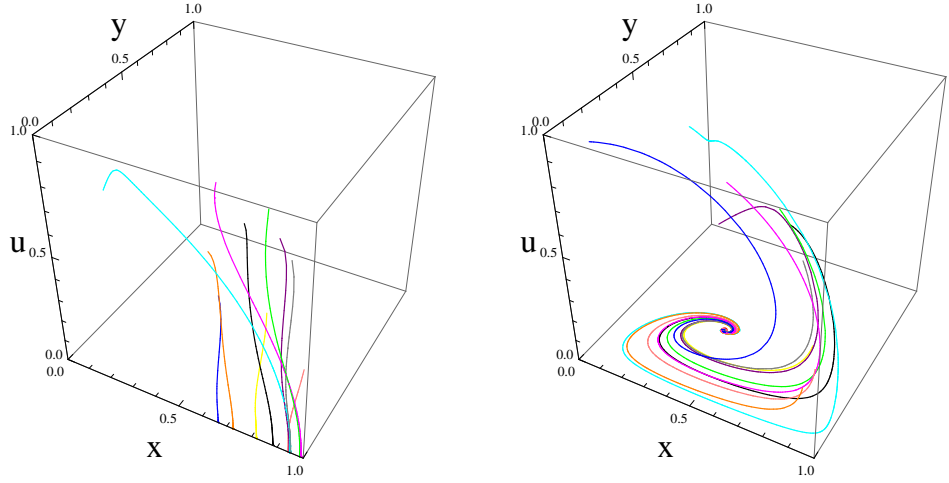}}
\caption{\label{fig1} The phase-space trajectories for the tracking attractors C3 and C5 with different initial conditions. The
left panel shows the critical point C3 with $(x_{c3},y_{c3},u_{c3})=(x,0,0)$ for the parameters $\epsilon=1$, $\lambda=15$ and $\gamma_b=2$.
The right panel shows the critical point C5 with $(x_{c5},y_{c5},u_{c5})=(1/\sqrt{6},1/\sqrt{6},0)$ for the parameters $\epsilon=1$, $\lambda=3$ and $\gamma_b=1$.}
\end{figure}

For the point C3, the eigenvalues are $\lambda_1=0$, $\lambda_2=3-\sqrt{\frac{3}{2}}\lambda x$ and $\lambda_3=-3$.
If $x<0$ and $\lambda>0$, $\lambda_2>0$. If $x>0$ and $\lambda^2\ge 6/x^2$, then $\lambda_2<0$. Since $\lambda_1=0$,
we need to study its stability further by using the center manifold theorem \cite{hkkhalil02}. For simplicity,
we take $\omega=0$, and the dynamical system (\ref{dynaeq1})--(\ref{dynaeq3}) reduces to the following system:
\begin{flalign}
\label{dynaeq1a}
&x'=\sqrt{\frac{3}{2}}\lambda y^2-3xy^2,\\
\label{dynaeq2a}
&y'=-\sqrt{\frac{3}{2}}\lambda xy +3y -3y^3.
\end{flalign}
To apply the center manifold theorem, we need to solve the equation
\begin{equation}
\label{dynaeq1c}
\frac{dh}{dx}\left(\sqrt{\frac{3}{2}}\lambda-3x\right)h^2+\left(\sqrt{\frac{3}{2}}\lambda x -3 +3h^2\right)h=0,
\end{equation}
with the initial condition $h(0)=h'(0)=0$. The solution is $y=h(x)=0$. Since the stability of the
dynamical system (\ref{dynaeq1a})-(\ref{dynaeq2a}) is the same as the system
$x'=0$ which is stable for the critical point, the point C3 is a stable point.
To illustrate its attractor behavior, we solve the dynamical system numerically
with different initial conditions for the parameters $\epsilon=1$, $\lambda=15$ and $\gamma_b=2$,
and the phase diagram is shown in the left panel of Fig \ref{fig1}.

For the point C4, the eigenvalues are $\lambda_1=\lambda_2=3/2$ and $\lambda_3=0$, so it is unstable.

For the point C5, the eigenvalues are
\begin{flalign}
&\lambda_1=\frac{3}{4} \left(-2 +\gamma_b-\frac{\sqrt{48 \epsilon\gamma_b^2 -24 \epsilon \gamma_b^3
+4  \lambda^2-20  \gamma_b \lambda^2+9  \gamma_b^2 \lambda^2}}{\left|{\lambda} \right|}\right),\nonumber \\
&\lambda_2=\frac{3}{4} \left(-2 +\gamma_b+\frac{\sqrt{48 \epsilon\gamma_b^2 -24 \epsilon \gamma_b^3 +4  \lambda^2-20  \gamma_b \lambda^2+9  \gamma_b^2 \lambda^2}}{\left|{\lambda} \right|}\right),\quad \nonumber\\
&\lambda_3=-\frac{3 \gamma_b}{2}.\nonumber
\end{flalign}
To keep the real parts of all three eigenvalues negative, we require that $0<\gamma_b<2$ and $\lambda^2>3\epsilon\gamma_b$.
The corresponding phase trajectories with different initial conditions for the parameters
$\epsilon=1$, $\lambda=3$ and $\gamma_b=1$ are shown in the right panel of Fig. \ref{fig1}.

\begin{figure}[htp]
\centerline{\includegraphics[width=0.8\textwidth]{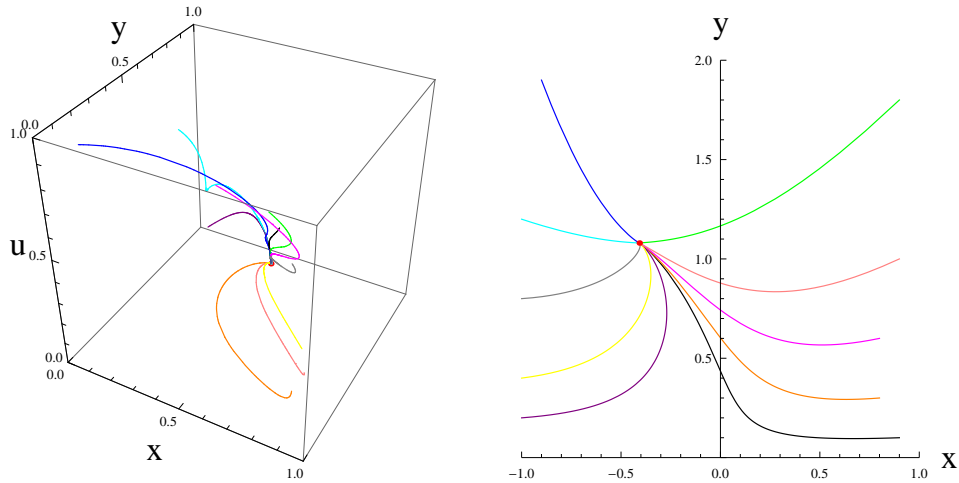}}
\caption{\label{fig2} The phase-space trajectories for the accelerating attractor C6 with different initial conditions.
The left panel is for the quintessence with $(x_{c6},y_{c6},u_{c6})=(1/\sqrt{6},\sqrt{5/6},0)$
for the parameters $\epsilon=1$, $\lambda=1$ and $\gamma_b=1$, and the right panel
is for the phantom with $(x_{c6},y_{c6},u_{c6})=(-1/\sqrt{6},\sqrt{7/6},0)$
for the parameters $\epsilon=-1$, $\omega=0$, $\lambda=1$ and $\gamma_b=1$.}
\end{figure}

For the point C6, the eigenvalues are $\lambda_1=-3+\lambda^2/(2\epsilon)$, $\lambda_2=-3\gamma_b+\lambda^2/\epsilon$ and   $\lambda_3=-\lambda^2/(2\epsilon)$.
For the quintessence case, $\epsilon=1$, so $\lambda_3<0$. The existence condition requires $\lambda^2<6\epsilon$, so $\lambda_1<0$.
If $\lambda^2<3\epsilon\gamma_b$, then $\lambda_2<0$. Therefore the stability condition
for the quintessence case is $\lambda^2<3\epsilon\gamma_b$. For the phantom case with $\omega\neq 0$, $\epsilon=-1$,
so $\lambda_3>0$ and the point is an unstable point. For the phantom case without the non-minimally
derivative coupling, $\omega=0$, the three dimensional system reduces to a two dimensional system,
the eigenvalue $\lambda_3$ is absent, $\lambda_1<0$ and $\lambda_2<0$, and the point is a stable point \cite{Hao:2003th}.
The corresponding phase trajectories with different initial conditions are shown in Fig. \ref{fig2}.
For the quintessence attractor, we take $\epsilon=1$, $\lambda=1$ and $\gamma_b=1$.
For the phantom attractor, we choose $\omega=0$, $\epsilon=-1$, $\lambda=1$ and $\gamma_b=1$.

For the point C7, we use the dynamical variable $v$ instead of $x$
to discuss the dynamical behavior and the eigenvalues are $\lambda_1=3\gamma_b-3$, $\lambda_2=\lambda_3=3\gamma_b/2>0$,
so it is an unstable point.

For the point C8, the dynamical variable $v$ instead of $x$
is used to discuss the dynamical behavior and
it was discussed in \cite{Gao:2010vr},
this de Sitter attractor is stable.

The properties of all the critical points are summarized in Table \ref{table1}.

\section{Discussion and Conclusions}

For the quintessence case with $\epsilon=1$ and $\omega=0$, in addition to the standard stable fixed points C5 and C6, we
also find the stable fixed point C3. The fixed points C3 and C5 are tracking solutions, and C6 gives the late time accelerating
solution with scalar field domination. For the phantom case with $\epsilon=-1$ and $\omega=0$, only the stable fixed point C6 exists.
For the case with non-minimally derivative coupling only, $\epsilon=0$ and $\omega\neq 0$, the dark matter solutions C1 and C4 are unstable,
only the de Sitter attractor exists. For the more general case with $\omega\neq 0$ and $\epsilon\neq 0$,
the stable fixed points C3, C5, and C6 exist only for
the quintessence field with $\epsilon=1$. C3 and C5 are tracking attractors and C6 is an accelerating attractor with dark energy
totally dominant if $\lambda^2<2$.

\begin{acknowledgements}
This research was supported in part by the Natural Science
Foundation of China under Grant Nos. 11175270 and 11475065,
the Program for New Century Excellent Talents in University under Grant No. NCET-12-0205
and the Fundamental Research Funds for the Central Universities under Grant No. 2013YQ055.
\end{acknowledgements}


\end{document}